# Possible bi-stable structures of pyrene-butanoic-acid-linked protein molecules adsorbed on graphene: Theoretical study

Yasuhiro Oishi[1], Motoharu Kitatani[1], and Koichi Kusakabe*[1],

Address: [1]Graduate School of Science, University of Hyogo, Kamigori, Hyogo 678-1297, Japan

Email: kusakabe@sci.u-hyogo.ac.jp

* Corresponding author

## Abstract

We theoretically analyze possible multiple conformations of protein molecules immobilized by 1-pyrenebutanoic-acid-succinimidyl-ester (PASE) linkers on graphene. The activation barrier between two bi-stable conformations exhibited by PASE is confirmed to be based on the steric hindrance effect between a hydrogen on the pyrene group and a hydrogen on the alkyl group of this molecule. Even after the protein is supplemented, this steric hindrance effect remains if the local structure of the linker consisting of an alkyl group and a pyrene group is maintained. Therefore, it is likely that the kinetic behavior of a protein immobilized with a single PASE linker exhibits an activation barrier-type energy surface between the bi-stable conformations on graphene. We discuss the expected protein sensors when this type of energy surface appears and provide a guideline for improving the sensitivity, especially as an oscillator-type biosensor.

## Keywords

PASE; protein; surface adsorption; biosensor; DFT;

## Introduction

Consideration of the atomic-scale motion of molecules based on nanoscience can lead to our better understanding of the behavior of target biomaterials and improve the sensitivity of specific dynamical systems, such as biosensors. In this study, we consider the behavior of proteins immobilized with linker molecules on graphene substrates. It is known that graphene may not easily adsorb proteins[1]. On the other hand, protein can be immobilized on graphene by using appropriate linker molecules, such as 1-pyrene-butanoic-acid succinimidyl ester (PASE). Actually, pyrene and its derivatives have been demonstrated to form stable bindings to carbon materials[2][3]. The properties and characteristics of these linker molecules are keys not only to the purpose of protein immobilization, but also to the behavior of the entire biosensor system.

In oscillator-based biosensors, further adsorption on the sensor, such as viruses using antigen/antibody reactions, may be detected via elastic-wave measurement. For this purpose, the antibody protein must first be immobilized on graphene. The antibody that specifically reacts with the target antigen is immobilized onto a graphene surface via the PASE linker. After that, the antigen is injected on sensor chips and specifically binds to the antibody. The antigen is then detected by observing the change in the vibrational frequency before and after the injection of the antigen.

Immobilization of protein using a PASE linker on carbon nanotube[1], graphite[4], and graphene[5] has been reported. The adsorption of PASE has been considered to mainly come from the pyrene fragment, which forms pi-pi stacking on these graphitic carbon materials[6]-[8].

The sensitivity of the oscillator-based sensor depends on the structure of the linker molecule. Therefore, understanding the adsorption mechanism of PASE linker on graphene and identifying characteristic conformations of adsorbed molecules are of great importance.

Recently, two of the present authors theoretically investigated the adsorption structure of PASE[9], revealing that PASE on graphene has a stable configuration (conformation 1) in a straight form on graphene. There is at least one metastable bent configuration (conformation 2) on the potential energy surface. Besides, the reaction barrier of the conformational change of the PASE was determined.

In this paper, we further examine a reaction pathway on the adiabatic potential energy surface that connects these conformations caused by the deformation of the PASE linker. The reaction pathway for the conformational change of the PASE itself was found to have the reaction activation barrier[9]. First, we identify that the origin of the barrier is the steric hindrance effect between two hydrogen atoms in the molecule.

A discussion on how the activation barrier between possible conformations of protein immobilized on graphene by linkers appears is also provided. We consider the adsorption of the linker molecule forming a bond with a protein, and derive the sufficient conditions for the adiabatic potential energy surface to have an activation barrier.

The reaction pathway having an activation barrier has an advantage in the detection of adsorbed molecules using elastic wave measurements. We explore the conditions under which this advantage can be expected for proteins immobilized on graphene. As a result, we propose a strategy for improving the accuracy of the sensing process in elastic wave measurement sensors through antigen/antibody reactions.

## PASE and PASE-derivatives on graphene

Among pyrene derivatives, PASE is widely used as a linker to connect carbon material and protein(Fig.1(a))[1][4][5]. The PASE linker and a certain protein can be connected through a dehydration condensation reaction so that the succinimide group is replaced by an amino group in a protein(Fig.1(b)). This dehydration-condensed linker with the protein may be used to immobilize proteins onto graphene. Since the connecting linker becomes a pyrene-butanoic-acid derivative (PBA), we will call them PBA-linked molecules. Adsorption of this molecule on graphene is known to happen by a mechanism similar to the adsorption of PASE on graphene[3,4].

The protein and pyrene are connected via an alkyl chain. The atomic configuration of this alkyl chain is believed not to be largely affected by the substitution of the succinimide group with an amino group. On the other hand, proteins alone do not readily form adsorbed structures on graphene[1]. Therefore, the pyrene moiety is bound to graphene, and at the same time, the protein is supplemented by a dehydration-condensed linker with the protein.

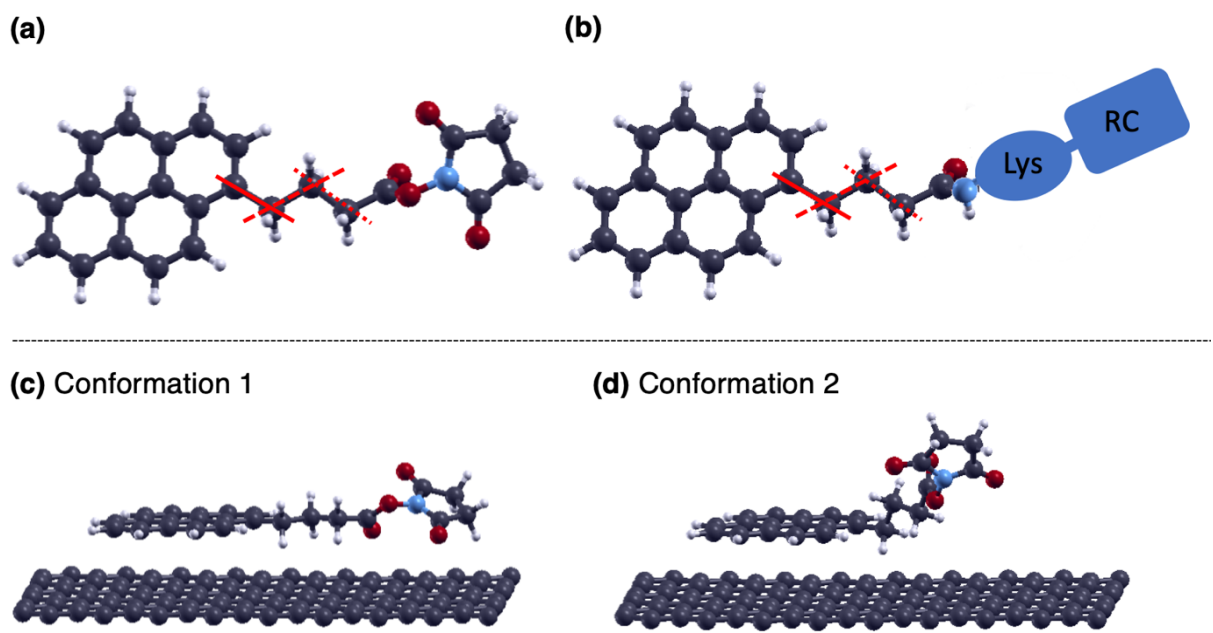

Figure 1(a-c): (a)Molecular structure of 1-pyrenebutanoic acid succinimidyl ester(PASE). The black, white, red, and blue ball represent the C, H, O, and N atoms., respectively. The solid, broken, and dotted red lines represent the rotation axis. Atomic geometry is visualized by Xcrysden[10] throughout this paper. (b) Schematic diagram of the dehydration-condensation structure of a hypothetical protein and PASE. Lys indicates lysine, and RC is the active center of the antigen/antibody reaction. (c,d) Structures of PASE linker on graphene for conformation 1 and 2, respectively[9].

As introduced, on graphene, PASE is found to be straight as the most stable conformation (conformation1 in Fig.1(c))[9]. On the potential energy surface, a curved conformation (conformation 2 in Fig.1(d)) exists. Our previous results suggested that conformation 2 is a metastable state. The relative stability of these conformations is known to vary depending on the surrounding environment. Actually, the presence of solutes consisting of water and proteins improves the stability of conformation 2.

# Mechanism of bi-stability in PASE on graphene

Conformations of PASE are interconverted by mutual transition when the molecule is twisted around one of the carbon-carbon single bonds in the alkyl chain. The rotational motion connecting conformation 1 and conformation 2 happens around the single bond indicated by the solid red line in Fig.1(a). The results of the nudged elastic band calculations show that the energy difference between the conformations is about 0.35 eV and the activation barrier on the path is about 0.55 eV(Fig.2) [9]. We will refer to the pathway on the potential energy surface with such an activation barrier as an activation-barrier type pathway.

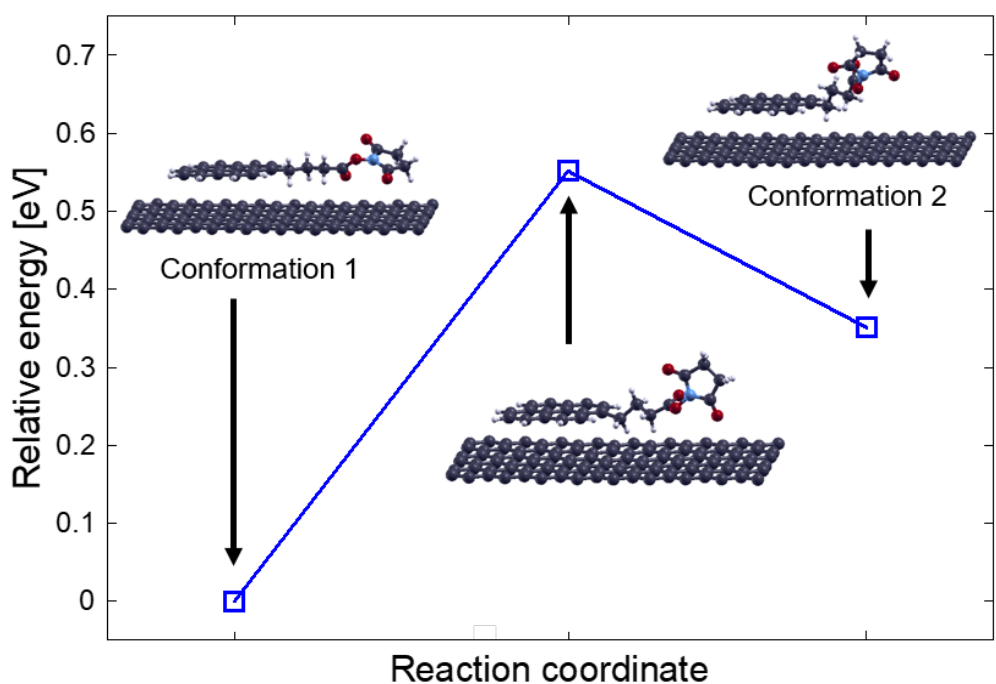

**Figure 2:** Relative energies of PASE on graphene for each conformation. The energy of conformation 1 is taken as an origin[9].

The pathway connecting conformations 1 and 2 appears as a torsion of the alkyl chain around the carbon-carbon single bond, as shown with the solid red line in Fig. 1(a). This is a kind of rotational motion. We checked the dihedral angles formed by the four carbon atoms connected by the red line in Fig. 3, and the values of dihedral angles for conformations 1, 2, and conformation at the activation barrier top are shown in Table 1. From these values, we can interpret the torsion as an approximate rotation. We call such rotations 0, π/3, and 2π/3 as an approximation, where conformation 1 corresponds to a 0 rotation, conformation 2 a 2π/3 rotation, and conformation at the activation barrier top a π/3 rotation, respectively.

To confirm the origin of the activation barrier that appears between conformation 1 and 2, we discuss the structural features of the activation barrier in detail. Figure 3 shows the structure at the activation barrier top. Two hydrogen atoms connected by the green line are characteristically close to each other. One hydrogen is on the pyrene skeleton, and the other is on the alkyl chain. The distance between these hydrogens is quite close, about 1.8 Å (Table 1). If we compare this to the distance between the same hydrogen pairs in conformation 1 and conformation 2, we can see that 1.8 Å is very short compared with that for the others (about 2.2 Å). Therefore, we conclude that such a local steric hindrance occurs at the activation barrier.

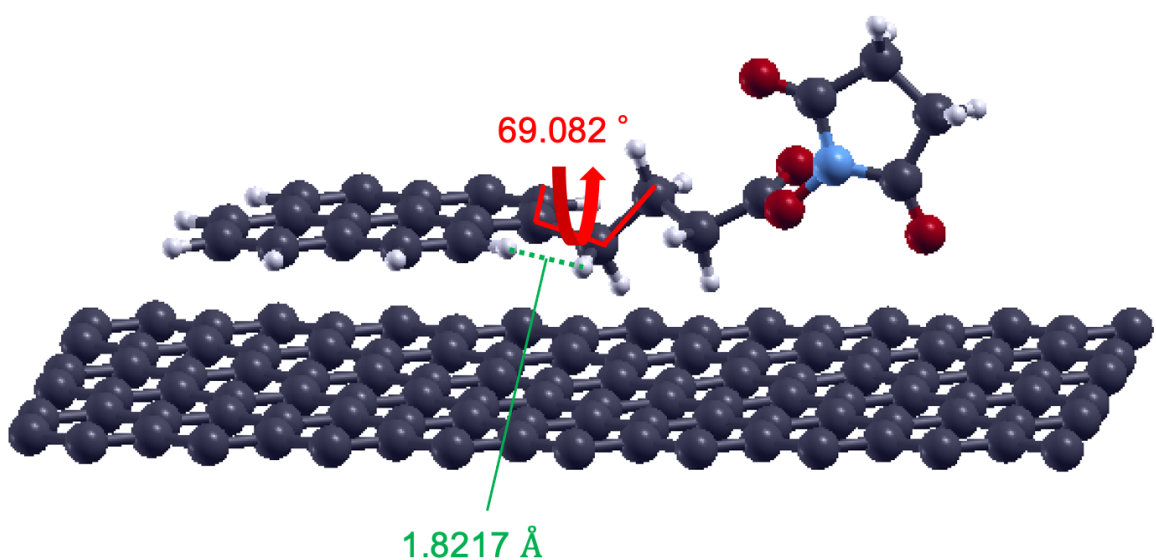

**Figure 3:** Structure of PASE on graphene at the activation barrier top[9].

**Table 1:** The dihedral angle formed by four carbon atoms connected by the red line in Figure 3, and the distance between hydrogen atoms connected by the green line in Figure 3, for conformation 1, 2, and conformation at the activation barrier top, respectively.

| Conformation | Dihedral angle [degree] | Distance between H atoms [Å] |
|---|---|---|
| Conformation 1 | 3.244 | 2.2443 |
| Activation barrier top | 69.082 | 1.8217 |
| Conformation 2 | 107.601 | 2.1785 |

This steric hindrance effect is analogous to the well-known steric hindrance effect of rotation around the C-C bond in ethane. Indeed, on the graphene surface, the possible conformations resulting from the rotation are effectively limited to conformation 2, due in part to the restricted direction of rotation. In this case, the activation barrier is created by the proximity of two hydrogen atoms.

There are other C-C single bonds in the alkyl chain, such as those shown by the dashed and dotted lines in Fig.1(a). Rotations around each bond can also happen. Thus, we expect similar steric hindrance effects to appear.

# Mechanical properties of PBA-linked molecular structure on graphene

Here, we discuss possible conformations of PBA-linked molecule. Would a similar activation-barrier type potential energy surface be expected to appear in the case of succinimide substitutions in proteins? To answer this question, we consider imide substitutions extensively. In principle, if the effect of substitution only provides a weak perturbation, the potential energy surface is only slightly changed and the barrier is maintained.

A substitution effect that does not effectively change the potential energy surface is isotopic substitution. In fact, the adiabatic potential surface defined for a hypothetical motion at the absolute zero degrees is not changed by a change in nuclear mass. Even in the density functional theory calculations, for example, the adiabatic potential energy surface, defined as the Born-Oppenheimer surface, is invariant for substitution between isotopes of carbon. This is an example of the weak perturbation.

In substitutions where the mass effect can be regarded as the predominant effect, it is safe to assume that similar steric hindrance effects appear when rotation around the alkyl chain occurs. Such an assumption may not be valid when substitutional binding of the linker to the protein causes some effects to the extent which rotation is not defined, such as a strong chemical bond change, a conformational change around the alkyl chain, or a distortion effect appearing in the adsorption structure of the linker itself. Those effects are not represented by mass substitution.

Thus, suppose that a substitution occurs to the extent which rotational motion around a C-C single bond in the alkyl chain is defined. Assume that this is a case that can be treated as a weak perturbation that manifests itself as a mass effect. Then, two (meta-)stable conformations would appear, one with the protein-containing part linearly lying on graphene, and the other with the conformation resulting from rotation at the alkyl chain. Rotational motion at the alkyl chain inevitably involves steric hindrance effects around the C-C bond. In particular, for motions corresponding to rotations around the C-C bond in PASE represented by the solid red line in Fig.1(a) and (b), the magnitude of the activation barrier is expected to be close to that in PASE. This is expected to be the case since a protein weakly bound to the carbon graphitic network can be supplemented onto graphene only by linkers as PASE, which has often been observed experimentally[1]. This type of proteins is weakly coupled to graphene. It can be concluded that except for bulky structures that allow multiple binding on graphene by several linkers, the above cases will be of a weakly perturbed nature when making a single link for protein capture.

## Discussion on strategy for improving biosensors

The vibrational properties of elastic waves, such as phase and amplitude, are known to be highly sensitive to the mass of adsorbed materials on graphene. Recently, a biosensor that detects target antigens by elastic wave measurement has been developed[11,12], using the method of immobilizing antibody protein on graphitic material by linker. The target antigen is introduced onto the sensor chips through antibody/antigen reaction.

Graphene has been considered as an ideal material for the platform of biosensors[13]-[23], benefiting from its excellent properties such as large detection area and high thermal conductivity[24]-[26]. Various biosensors have already been developed, many of which require immobilization of antibody protein on graphene surface. Therefore, protein capture by linkers is used in a relatively large number of sensors.

In the elastic wave measurement system, elastic waves are generated by laser irradiation and their responses are observed. The elastic wave is excited by a pump light, and the response of the substrate is observed via the probe light. The refractive index of metallic substrates such as graphene may change due to the arrival of elastic waves, and this change can be observed by detecting the reflected light of the probe light. For such laser-based biosensors, graphene's conductivity can provide a unique advantage. In fact, the conductivity of multi-layered graphene is highly anisotropic. Therefore, the use of multilayer graphene avoids the burning of biomaterials by lasers irradiated from the backside of the substrate. The present study is concerned with the mechanical deformation of the linker molecule. The mechanical properties of the rotational motion are almost the same whether modeled in single-layer graphene or in multilayer graphene.

Here, we will see how the sensitivity as a sensor can be improved when there is such rotational motion of the linked molecules adsorbed on graphene. Antigen/antibody reactions occur at the active centre in the protein portion. Therefore, the reaction at this active centre should be more activated at the stage of contact with the antigen. In a relatively large number of cases, specimens such as viruses are collected in a captured state in a solvent. The specimen is introduced onto a graphene substrate linked with a protein prepared as a sensor. As already mentioned, the stability of conformation 2 is improved in a solution. Therefore, the rotational motion at the linker portion should be more active in a solution, and this will lead to an increase in the probability that the active centre of the antigen/antibody reaction reacts with the antigen. Therefore, it is desirable to select the solvent and temperature conditions so that the probability of appearance of conformation 2 increases.

On the other hand, at the time of sensing, the proteins and sample antigens bound by the linker should be strongly bound on the graphene surface. This makes it easier to observe the increased mass by generating a reflected wave with a strong amplitude relative to the incident elastic wave. Therefore, it is advisable to adjust the protein portion containing the supplemented antibody to be on the graphene surface corresponding to conformation 1 by evaporating the solvent or by changing the temperature conditions.

Here, we mention another merit that a pyrene derivative has for another type of biosensor. In immobilizing proteins on graphene, a pyrene derivative has another merit: non-covalent functionalization, i.e., weaker in bonding with graphene than covalent one for functionalization. Actually, widely employed covalent functionalization may not be suitable for a special purpose[27]. This approach is known to have an undesirable effect of disturbing the electronic properties of graphene[15]. On the other hand, in non-covalent functionalization, the electronic properties of graphene can be preserved[28]-[30]. Therefore, the PASE linker and its properties may be relevant for another type of the electronic sensing strategy, such as field effect transistor-based biosensors.

## Conclusion

We discussed how an activation barrier appears on the potential energy surface in PASE adsorbed on graphene. The conformational change between two bi-stable PASE structures can be regarded as rotational motion around a C-C bond in the alkyl chain. An origin of an activation barrier is the steric hindrance coming from the proximity of a hydrogen on pyrene to that on alkyl group.

We also considered the state after a protein is supplemented, and discussed how a similar potential energy surface to that in PASE is expected. If the supplementation of protein can be regarded as a weak perturbation, the rotational motion is expected to remain and the activation barrier is maintained. Then, the probability of the appearance of conformation 2 relative to conformation 1 and vice versa can be adjusted by arranging the environment such as temperature and solvent.

We provided a guideline for an improvement of biosensing device. In oscillator-type biosensor, by arranging an appropriate surrounding environment, an improvement of the sensitivity is expected.

# Computational method

The density functional theory (DFT) calculation determined stable atomic configurations of conformation 1 and 2 of PASE on graphene. In the DFT calculation realized by the PWscf code of Quantum ESPRESSO[31]-[33], the DFT-D3 correlation[34] together with PBEsol functional[35] for the exchange-correlation functional described van der Waals interaction between graphene and pyrene fragment in PASE. Ultrasoft pseudopotentials[36] with the energy cutoff of 35 (350) Ry for the expansion of wavefunction (charge density) described the electron-nuclear interaction. Using the $2\times4\times1$ ***k***-point mesh of Monkhorst-Pack[37], the Brillouin zone sampling was safely performed. Additional details of calculation conditions may be found in a previous paper[9].

After the structural optimization calculation of the (meta)stable conformations until the Hellman-Feynman force acting on each atom was less than $10^{-6}$ Ry/Bohr, the nudged elastic band method determined the minimum energy pathway. Intermediate images created by linear interpolation between conformation 1 and 2 were optimized so that each image had lower energy while adequate spaces between neighboring images on the energy surface were kept.

To explore the origin of the activation barrier, we analyzed the structure of the saddle point on the minimum energy pathway. Having conformation 1, 2, and conformation at the saddle point, we measured the dihedral angles and the distance between two hydrogen atoms using Xcrysden[10].

# Acknowledgements

We are grateful to Prof. H. Ogi, Prof. M. Ohtani, Prof. N. Nakamura, Prof. K. Tanigaki, Prof. S. Hagiwara, and Prof. K. Nakajima for fruitful discussions. The calculations were done at the computer centers of Kyushu University and ISSP, University of Tokyo.

# Funding

This work is partly supported by JSPS KAKENHI Grant No. JP22K04864.